\documentclass[letterpaper,10pt,conference]{ieeeconf}
\IEEEoverridecommandlockouts
\usepackage{amsmath}
\usepackage{amssymb}
\usepackage{bm}
\usepackage{graphicx}
\usepackage{booktabs}
\usepackage{placeins}

\title{\LARGE \bf
Preference-Adaptive Control in Autonomous Driving
}
\author{
Zhongyu Mo$^{1}$,
Shanting Wang$^{1}$,
Huaijin Hu$^{1}$,
and Andreas A. Malikopoulos$^{2}$,
\textit{Senior Member, IEEE}
\thanks{This research was supported in part by NSF under Grants CNS-2401007, CMMI-2348381, IIS-2415478, and in part by MathWorks.}\thanks{$^{1}$Zhongyu Mo, Shanting Wang, and Huaijin Hu are with the Systems
Engineering Program, Cornell University, Ithaca, NY 14853 USA (emails:
\{zm73, sw997, hh833\}@cornell.edu).}\thanks{$^{2}$Andreas A. Malikopoulos is with the Applied Mathematics, Systems Engineering, Robotics, Mechanical Engineering, Electrical and Computer Engineering, and Civil \& Environmental Engineering, Cornell University, Ithaca, NY, USA. (email: \texttt{amaliko@cornell.edu})}}

\begin{document}
\maketitle
\thispagestyle{empty}
\pagestyle{empty}

\begin{abstract}
In this paper, we present a preference-adaptive receding-horizon control framework for autonomous driving that accounts for passenger preferences and motion-sickness susceptibility. We formulate a finite-horizon optimal control problem with adaptive weights for speed, acceleration comfort, and motion sickness, while maintaining a fixed weight for collision risk. We predict motion sickness online using an individualized model on the Motion Illness Symptoms Classification (MISC) scale and update the preference weights offline from emotion-derived pairwise comparisons using Bayesian inference. A deterministic safety supervisor checks the planned trajectory and modifies the control command when necessary. We evaluate the framework using three simulated passenger profiles under three motion-sickness susceptibility levels. The learned weights yield distinct closed-loop behaviors, and the mean evaluation emotion score improves in seven of nine scenarios. No collisions occur in the full-method experiments, while the safety supervisor intervenes in 1.621\% of the learning frames.
\end{abstract}

\section{Introduction}

\begin{figure*}[t]
\vspace*{5pt}
\centering
\includegraphics[width=2\columnwidth]
{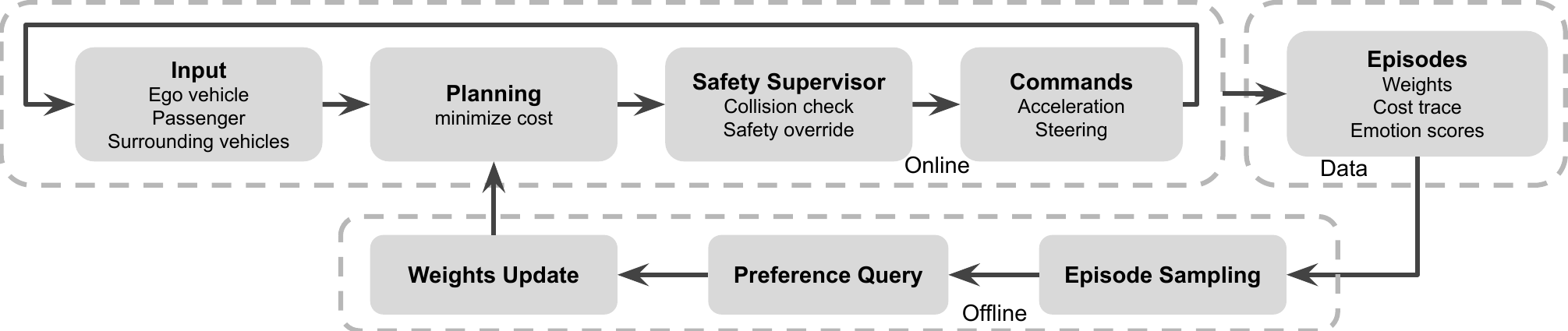}
\caption{Preference-adaptive control architecture. Online, ego-vehicle, passenger, and surrounding-vehicle states enter finite-horizon planning, after which the safety supervisor validates or overrides the planned command. Closed-loop episodes provide cost and emotion data for offline episode selection, preference queries, and weight updates, and the updated weights are returned to the planner.}
\label{fig:Pipeline}
\vspace{-5pt}
\end{figure*}

Autonomous vehicles can improve safety and mobility, but passengers
may differ substantially in their desired tradeoff between speed and comfort and in their susceptibility to motion sickness. Thus, a fixed
control objective may not adequately represent an individual
passenger's preferences.

Personalized vehicle control has been studied using inverse
reinforcement learning~\cite{Zhao2022}, trajectory comparisons and
preference learning~\cite{Wen2024,Ellis2024}, and Bayesian
optimization for controller adaptation
~\cite{Le2023ACC,le2025controller}. These approaches provide
mechanisms for adapting vehicle behavior, but they do not directly
account for individualized, history-dependent motion sickness when
learning passenger preferences.

Motion sickness depends on the accumulated interaction between
vestibular and visual information and cannot, in general, be
represented by instantaneous acceleration alone. Sensory-conflict
models have been developed for motion-sickness prediction
~\cite{Wada2020,Kotian2024}, while Kotian et al.
~\cite{Kotian2023,Kotian2025} introduced individualized accumulation
models based on the framework of Oman~\cite{Oman1990}. Recent studies
have further investigated motion-sickness prediction and mitigation
for vehicle passengers~\cite{Wada2021,Tian2025,Emond2026,Tang2026}.
However, these models do not address how to infer an individual
passenger's tradeoff among motion sickness, speed, and acceleration
comfort from passenger feedback. Emotion recognition provides one
possible feedback channel for passenger acceptance~\cite{Marceddu2021}.

We address this gap with a preference-adaptive receding-horizon
control framework that learns the relative importance assigned by a
passenger to speed, acceleration comfort, and motion sickness. Because
motion sickness evolves through history-dependent accumulation, we
incorporate its individualized prediction directly into finite-horizon
planning. The passenger-dependent weights are updated offline from
emotion-derived pairwise comparisons using Bayesian inference and
Markov chain Monte Carlo (MCMC) sampling as in~\cite{Wen2024}. During
online operation, an independent deterministic safety supervisor
modifies the planned command when short-horizon collision or
road-boundary checks fail
~\cite{Schwarz2014,Hou2014,ChalakiCBF2022,wang2025visiopath}.

The contributions of this paper are the following. First, we formulate
a passenger-adaptive finite-horizon optimal control problem that
incorporates individualized, history-dependent motion-sickness
prediction with speed and acceleration comfort. Second, we adapt
pairwise Bayesian preference learning to update the passenger-dependent
weights from realized closed-loop episodes and use these weights
directly in the receding-horizon controller. Finally, we evaluate the
framework for three simulated passenger profiles under three levels of
motion-sickness susceptibility. The results yield distinct
passenger-dependent closed-loop behaviors and improve the mean
evaluation emotion in seven of the nine scenarios while exposing
limitations associated with threshold-sensitive preferences.

The remainder of the paper is organized as follows. In Section~II, we present the vehicle and passenger models and formulate the finite-horizon optimal control problem. In Section~III, we present the preference-learning framework, online planning approach, and safety supervisor. In Section~IV, we present the simulation setup and results. Finally, we provide concluding remarks in Section~V.

\section{Problem Formulation}

We consider an autonomous ego vehicle carrying one passenger on a
multi-lane highway in the presence of surrounding vehicles. Our
objective is to account for passenger preferences for speed,
acceleration comfort, and motion sickness while satisfying vehicle
constraints and accounting for collision risk.

\textit{Assumption 1.}
At each control step, the ego vehicle has access to its state and the
states of the surrounding vehicles considered by the planner.

This assumption separates preference adaptation from state estimation.
These states can be obtained from onboard localization and perception;
in our simulations, they are provided directly by the environment.

\subsection{Vehicle Dynamics and Constraints}

Let
\[
s_k=[x_k,y_k,\psi_k,v_k,\delta_k]^T,\qquad
u_k=[a_{x,k},\delta_k^{cmd}]^T
\]
denote the ego-vehicle state and control input, respectively. Here,
$(x_k,y_k)$ are the longitudinal and lateral coordinates
of the vehicle center of mass, $\psi_k$ is the yaw angle, $v_k$ is the
vehicle speed, and $\delta_k$ is the applied steering angle. The
control input consists of the longitudinal acceleration $a_{x,k}$ and
commanded steering angle $\delta_k^{cmd}$. 

\begin{figure}[h]
\centering
\includegraphics[width=0.68\columnwidth]{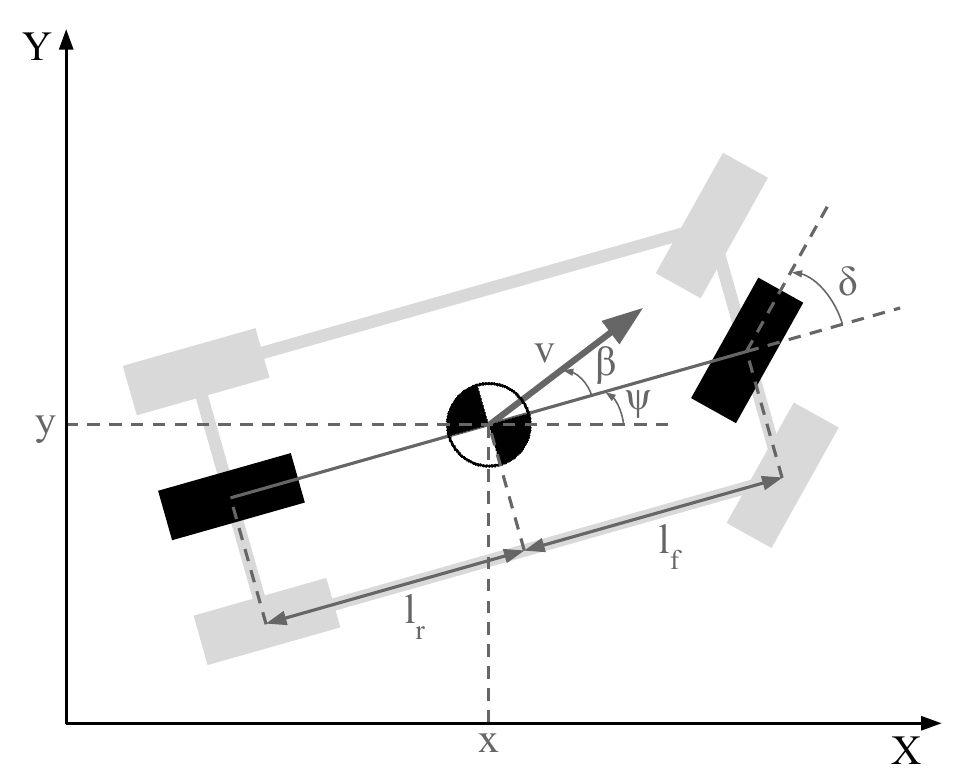}
\caption{Kinematic bicycle model and ego-vehicle notation. The vehicle center is located at $(x_k,y_k)$ with speed $v_k$, yaw angle $\psi_k$, steering angle $\delta_k$, slip angle $\beta_k$, and axle distances $l_f$ and $l_r$.}
\label{fig:kinematic_bicycle}
\end{figure}

We use the kinematic
bicycle model in~\cite{kong2015}, illustrated in
Fig.~\ref{fig:kinematic_bicycle}, and discretize it with a midpoint method.

\begin{equation*}
\begin{aligned}
\beta_k &=
\tan^{-1}\!\left(\frac{l_r}{l_f+l_r}\tan\delta_k^{cmd}\right),\\
v_{k+\frac12} &=
\max\!\left\{0,v_k+\frac{\Delta t}{2}a_{x,k}\right\},\\
\psi_{k+\frac12}
&=
\psi_k+
\frac{\Delta t}{2}
\frac{v_k\cos\beta_k\tan\delta_k^{cmd}}{l_f+l_r}.
\end{aligned}
\end{equation*}
The discrete dynamics are
\begin{equation}
\begin{aligned}
x_{k+1}
&=x_k+\Delta t\,v_{k+\frac12}
\cos(\psi_{k+\frac12}+\beta_k),\\
y_{k+1}
&=y_k+\Delta t\,v_{k+\frac12}
\sin(\psi_{k+\frac12}+\beta_k),\\
\psi_{k+1}
&=\psi_k+\Delta t\,
\frac{v_{k+\frac12}\cos\beta_k\tan\delta_k^{cmd}}
{l_f+l_r},\\
v_{k+1}
&=\max\{0,v_k+\Delta t\,a_{x,k}\},\\
\delta_{k+1}
&=\delta_k^{cmd}.
\end{aligned}
\label{eq:kinematic_bicycle_discrete}
\end{equation}
The corresponding yaw rate and lateral acceleration are
\begin{equation}
\begin{aligned}
\omega_{z,k}
&=
\frac{v_{k+\frac12}\cos\beta_k\tan\delta_k^{cmd}}
{l_f+l_r},\\
a_{y,k}
&=v_{k+1}\omega_{z,k}.
\end{aligned}
\label{eq:yaw_lateral_acceleration}
\end{equation}

We impose the input constraints
\begin{equation}
\begin{gathered}
|\delta_k^{cmd}-\delta_k|
\le \dot{\delta}_{\max}\Delta t,\quad
|\delta_k^{cmd}|
\le \delta_{\max},\\
a_{x,\min}
\le a_{x,k}\le a_{x,\max}.
\end{gathered}
\label{eq:input_constraints}
\end{equation}
and the motion constraints
\begin{equation}
\begin{gathered}
0\le v_k\le v_{\max},\quad
|\psi_k|\le\psi_{\max},\quad
|a_{y,k}|\le a_{y,\max}.
\end{gathered}
\label{eq:motion_constraints}
\end{equation}

For a vehicle of length $l_L$ and width $l_W$, define its lateral
extent as
\[
h_y(\psi_k)
=
\frac{l_L}{2}|\sin\psi_k|
+
\frac{l_W}{2}|\cos\psi_k|.
\]
The vehicle footprint must satisfy
\begin{equation}
\begin{gathered}
y_{\min,\mathrm{road}}
\le y_k-h_y(\psi_k),\quad
y_k+h_y(\psi_k)
\le y_{\max,\mathrm{road}},
\end{gathered}
\label{eq:road_constraints}
\end{equation}
and we impose the additional center-position constraint
\begin{equation}
y_{\min,\mathrm{ego}}
\le y_k\le
y_{\max,\mathrm{ego}}.
\label{eq:ego_center_constraint}
\end{equation}
The latter provides an additional planning margin from the road boundaries.

Each
surrounding vehicle $i$ is represented by its lane, longitudinal
position $x_{i,k}$, speed $v_{i,k}$, and desired speed $v_i^{des}$.

\subsection{Passenger State and Preference Signal}

We characterize motion sickness using the Motion Illness Symptoms
Classification (MISC) scale, where $M=0$ represents no symptoms and
$M=10$ vomiting. To predict $M_k$ on this scale, we adopt the
individualized model of Kotian et al.~\cite{Kotian2025}. The model
first generates sensory conflict using the group-averaged Subjective
Vertical Conflict (SVC) model from Wada~\cite{Wada2020}, and then
accumulates this conflict using an individualized two-parameter
accumulation model (AM2) based on Oman~\cite{Oman1990}.

Let $\chi_k=(\xi_k,z_{1,k},z_{2,k})$ collect the internal model states,
where $\xi_k$ is the SVC state and $z_{1,k},z_{2,k}$ are the AM2
states, and let $\theta=(\theta_T,\theta_K)$ denote the individual
susceptibility parameters. We write the discretized passenger model
compactly as
\begin{equation}
(\chi_{k+1},M_{k+1})
=
F_M^{\Delta t}
(\chi_k,a_k,\omega_k,g_k;\theta),
\label{eq:passenger_state}
\end{equation}
where $a_k,\omega_k\in\mathbb{R}^3$ are the vehicle linear acceleration and
angular velocity, respectively, and
$g_k\in\{\mathrm{internal},\mathrm{external},\mathrm{closed}\}$
denotes the passenger vision mode. The mapping $F_M^{\Delta t}$
denotes our discretized implementation of the individualized
motion-sickness model in~\cite{Kotian2025}.

\textit{Assumption 2.}
The susceptibility parameters $\theta_K$ and $\theta_T$ are known for
each passenger.

These parameters can be identified independently using the
method in~\cite{Kotian2025}; their identification is outside the scope of this paper.

\textit{Assumption 3.}
The passenger vision mode $g_k$ is available at each control step, and
the instantaneous emotion score $e_k$ is measurable during each episode.

The vision mode is required for motion-sickness prediction, while the
emotion score provides the preference feedback used for updating weights offline. These quantities may be obtained from occupant-monitoring
systems~\cite{Marceddu2021}; in our simulations, they are generated
directly from the passenger profiles in Section~IV.

\subsection{Collision-Risk Cost}

\begin{figure}[h]
\centering
\includegraphics[width=0.92\columnwidth]
{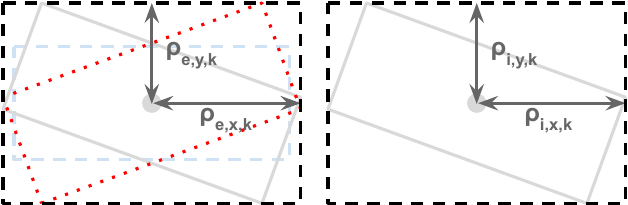}
\vspace{-1em}
\caption{Left: the black dashed rectangle encloses the ego-vehicle
footprints from the start of timestep $k$ (gray solid) to the start of
$k+1$ (red dotted), and is aligned with the midpoint ego heading at
$k+\frac{1}{2}$ (blue dashed). Right: encloses surrounding vehicle $i$ in the same ego-midpoint-heading frame.}
\label{fig:TTC}
\end{figure}

We use time to collision (TTC) as a soft measure of collision risk.
Motivated by prior work using 2D TTC as in~\cite{Schwarz2014} and ellipse
approximations as in~\cite{Hou2014}, we conservatively enclose the collision
region between the ego vehicle and surrounding vehicle $i$ by a rotated ellipse,
aligned with the midpoint ego heading and enlarged to contain the rectangle-overlap
region for all ego orientations between the start and end headings of prediction
step $k$, as illustrated in Fig.~\ref{fig:TTC}.

Let
\[
\begin{aligned}
d_{x,i,k}(t)&=\Delta x_{i,k}+v^{\mathrm{rel}}_{x,i,k}t,\\
d_{y,i,k}(t)&=\Delta y_{i,k}+v^{\mathrm{rel}}_{y,i,k}t,\\
A_{i,k} &= \sqrt{2}\cdot(\rho_{e,x,k}+\rho_{i,x,k}),\\
B_{i,k} &= \sqrt{2}\cdot(\rho_{e,y,k}+\rho_{i,y,k}),
\end{aligned}
\]
where $\Delta x_{i,k},\Delta y_{i,k}$ and
$v^{\mathrm{rel}}_{x,i,k},v^{\mathrm{rel}}_{y,i,k}$ are the relative
position and velocity components in the ellipse frame. $\rho_{e,x,k}$ and $\rho_{e,y,k}$ denote the longitudinal and lateral half-extents of the ego enclosing rectangle, and $\rho_{i,x,k}$ and $\rho_{i,y,k}$ denote the corresponding half-extents of vehicle $i$, all in the midpoint-heading frame. $A_{i,k}$ and $B_{i,k}$ are the longitudinal and lateral semi-axis lengths of the conservative collision ellipse for the ego vehicle and vehicle $i$. Then
$TTC_{i,k}$ is the smallest nonnegative solution of
\begin{equation}
\left(
\frac{d_{x,i,k}(TTC_{i,k})}{A_{i,k}}
\right)^2
+
\left(
\frac{d_{y,i,k}(TTC_{i,k})}{B_{i,k}}
\right)^2
=1,
\label{eq:TTC}
\end{equation}
We set
$TTC_{i,k}=0$ if vehicle $i$ is already inside the ellipse and
$TTC_{i,k}=\infty$ if no nonnegative solution exists.

With $TTC_k=\min_i TTC_{i,k}$, we define
\begin{equation}
J_{TTC}(k)=
\left[
1-\min\left\{
\frac{TTC_k}{TTC_{th}},1
\right\}
\right]^2.
\label{eq:ttc_cost}
\end{equation}
Thus, $J_{TTC}=0$ for $TTC_k\ge TTC_{th}$ and increases as the
predicted time to collision decreases. Since TTC is a soft penalty,
Section~III introduces an independent safety supervisor.

\subsection{Optimization Problem}

Let $w=[w_v,w_M,w_a]^T$ denote the passenger preference vector. We define the
passenger-dependent stage cost as
\[
J_p(k)
=
w_vJ_v(k)+w_MJ_M(k)+w_aJ_a(k).
\]
The fixed weights $\lambda_p,\lambda_s>0$, with
$\lambda_p+\lambda_s=1$, balance passenger preference and collision
risk. We keep $\lambda_s$ fixed so that learning changes only the
relative importance of the passenger-dependent objectives.

\textit{Problem 1.}
Given the current ego, passenger, and surrounding-vehicle states and
the preference vector $w$, derive $\{u_k\}_{k=0}^{H-1}$ that solves
\begin{equation}
\min_{\{u_k\}_{k=0}^{H-1}}
\sum_{k=0}^{H-1}
\left[
\lambda_pJ_p(k)
+
\lambda_sJ_{TTC}(k)
\right]
\label{eq:objective}
\end{equation}
subject to
\eqref{eq:kinematic_bicycle_discrete},
\eqref{eq:passenger_state}, and
\eqref{eq:input_constraints}--\eqref{eq:ego_center_constraint}.
The costs $J_v$, $J_M$, and $J_a$ are defined in Section~III-A.

\section{Preference-Adaptive Receding-Horizon Control}

In this section, we present the passenger-dependent costs, preference
learning, receding-horizon implementation, and safety supervisor. The overall pipeline is depicted in Fig.~\ref{fig:Pipeline}.

\subsection{Passenger Preference Costs}

We define costs for motion sickness, speed, and acceleration comfort.
Let $M_0<M_1$ denote the onset and end of the rapid
motion-sickness penalty transition, let $M_{\max}$ denote the maximum
MISC value, and let $\alpha_M\in[0,1]$ balance the rapid transition
and the gradual severity term. Define
\[
h(z)=6z^5-15z^4+10z^3
\]
and
\[
\widetilde M_{k+1}
=
\begin{cases}
0, & M_{k+1}\le M_0,\\
\dfrac{M_{k+1}-M_0}{M_1-M_0},
& M_0<M_{k+1}<M_1,\\
1, & M_{k+1}\ge M_1.
\end{cases}
\]
The motion-sickness cost is
\begin{equation}
J_M(k)
=
\alpha_M h(\widetilde M_{k+1})
+
(1-\alpha_M)\frac{M_{k+1}}{M_{\max}}.
\label{eq:misc_cost}
\end{equation}
Thus, $J_M$ remains small before the rapid-transition region and
increases sharply as $M$ approaches the noticeable-discomfort
threshold.

The remaining costs are
\begin{equation}
\begin{aligned}
J_v(k)
&=
\left(
\frac{v_{\max}-v_{k+1}}{v_{\max}}
\right)^2,\\
J_a(k)
&=
\frac{1}{2}
\left[
\left(\frac{a_{x,k}}{a_{x,\lim}}\right)^2
+
\left(\frac{a_{y,k}}{a_{y,\max}}\right)^2
\right],
\end{aligned}
\label{eq:preference_costs}
\end{equation}
where $a_{x,\lim}=a_{x,\max}$ during acceleration and
$a_{x,\lim}=|a_{x,\min}|$ during braking. The weights in $w$ therefore
determine the tradeoff among speed, motion sickness, and acceleration
comfort.

We use the same cost for all
passengers; individual differences enter through the susceptibility
parameters and learned weight $w_M$.

\subsection{Bayesian Preference Learning}

We next use the realized episode costs and emotion scores to update the passenger preference weights.

Let $\tau$ denote a closed-loop episode. We define its reward-feature
vector as
\begin{equation}
r(\tau)
=
[-J_v^\tau,-J_M^\tau,-J_a^\tau]^T,
\label{eq:reward_features}
\end{equation}
where each cost is accumulated over the episode.

\textit{Assumption 4.}
For two episodes, the passenger prefers the one
with higher episode-mean emotion score.

This assumption converts the measured affective response into the
binary comparisons required for preference learning without requiring
the passenger to recall and compare complete trips. In our simulations,
the emotion signal is generated by the passenger profiles in
Section~IV. For a collision-terminated episode, the accumulated
reward features are scaled to the nominal episode duration before comparison.

Following~\cite{Wen2024}, define
$p_{ab}(w)=p(\tau_a>\tau_b\mid w)$. Then
\begin{equation}
p_{ab}(w)
=
\frac{\exp(w^Tr(\tau_a))}
{\exp(w^Tr(\tau_a))+\exp(w^Tr(\tau_b))}.
\label{eq:wen16}
\end{equation}
Since $p_{ba}(w)=1-p_{ab}(w)$, we select informative episode pairs
using
\begin{equation}
\max_{(\tau_a,\tau_b)}
\min
\left\{
\mathbb{E}_w[1-p_{ab}(w)],
\mathbb{E}_w[p_{ab}(w)]
\right\}.
\label{eq:wen17}
\end{equation}
Let $N_q$ denote the number of preference labels accumulated so far. Given preference labels $q_1,\ldots,q_{N_q}$, the posterior satisfies
\begin{equation}
p(w\mid q_{1:N_q})
\propto
p(w)\prod_{m=1}^{N_q}p(q_m\mid w).
\label{eq:wen18}
\end{equation}

\textit{Assumption 5.}
The passenger's underlying preference remains unchanged during each learning
experiment.

This assumption is consistent with the fixed-parameter Bayesian model.
Time-varying preferences would require a dynamic preference model or
discounting of previous comparisons.

We update the posterior estimate of $w$ after every $L$ new eligible episodes. Eligible
pairs contain at least one new episode and satisfy
$\|\Delta r\|>10^{-8}$. We rank them using~\eqref{eq:wen17}, retain
up to $L$ comparisons, assign labels using Assumption~4, and discard
ties. To enforce positive weights that sum to one, we use $w=\operatorname{softmax}(\eta)$,
where $\eta$ is unconstrained. We place independent standard-normal
priors on its components. We sample the posterior using MCMC and use the mean normalized sample as the
preference vector for subsequent episodes.

\subsection{Receding-Horizon Implementation}

At each control step, we solve Problem~1 using Sequential Least-Squares
Programming (SLSQP)~\cite{daza2023}. We propagate the ego and passenger
states using~\eqref{eq:kinematic_bicycle_discrete} and
\eqref{eq:passenger_state}, respectively, and predict surrounding
vehicles using the Intelligent Driver Model (IDM).

\textit{Assumption 6.}
The passenger-specific vision policy is known to the controller.

This assumption avoids introducing a separate model of future gaze
behavior. In our simulations, the known vision policy is evaluated at each
predicted step using the corresponding predicted MISC state.

We enforce
\eqref{eq:input_constraints}--\eqref{eq:ego_center_constraint}
throughout the prediction horizon and execute only the first control
command. If SLSQP does not return a feasible solution, we use a
fallback sequence with zero longitudinal acceleration and steering
toward the global-forward heading. Otherwise, we shift the previous
solution and append a neutral command to initialize the next
optimization. Because this fallback is generated directly rather than by constrained
optimization, satisfaction of the nominal planning constraints is not
guaranteed. In either case, the safety supervisor evaluates the
planned trajectory before execution.

\subsection{Safety Supervisor}

Since $J_{TTC}$ is a soft collision-risk penalty, we independently
evaluate the planned trajectory over a short horizon $H_S$. We call a trajectory \emph{admissible} if, throughout \(H_S\), the predicted ego footprint and center remain within the prescribed road bounds and no predicted relative trajectory enters a conservative pairwise collision envelope. This condition depends on predicted surrounding-vehicle motion and therefore does not constitute a formal safety guarantee.

If the planned sequence $u^{plan}=\{u_k^{plan}\}_{k=0}^{H-1}$ is not admissible, the supervisor
searches a finite set $U^S$ of constant commands
$u^*=[a_x^*,\delta^{cmd,*}]^T$. Let
$U^{adm}\subseteq U^S$ denote the admissible candidates. For each
$u^*\in U^{adm}$, we define
\begin{equation}
\begin{aligned}
J_S(u^*)
=
\frac{1}{H_S}\sum_{k=1}^{H_S}
\Bigg[
&\left(\frac{v_k-v_S}{v_{\max}}\right)^2\\
&+
\left(\frac{\psi_k-\psi_S}{\psi_{\max}}\right)^2
\Bigg],
\end{aligned}
\label{eq:supervisor_cost}
\end{equation}
where $v_S$ and $\psi_S$ are the speed and heading when the supervisor
is triggered. The executed command is
\begin{equation}
u^{exec}
=
\begin{cases}
u_0^{plan},
& u^{plan}\text{ admissible},\\[0.3em]
\displaystyle
\arg\min_{u^*\in U^{adm}}J_S(u^*),
& U^{adm}\neq\emptyset,\\[0.6em]
\displaystyle
\arg\max_{u^*\in U^S}t_{vio}(u^*),
& U^{adm}=\emptyset,
\end{cases}
\label{eq:safety_supervisor}
\end{equation}
where $t_{vio}(u^*)$ is the first predicted violation time. If no
admissible candidate exists, the supervisor therefore selects the
command that delays the predicted violation for the longest time.

\section{Simulation Framework and Evaluation Protocol}

\subsection{Simulation Environment}

We consider a straight three-lane highway with lane width $3.6$~m
and $18$ surrounding vehicles governed by IDM, with desired speeds in the range \(12\)–\(21\) m/s. The ego
vehicle starts in the center lane at $16$~m/s. We impose
$0\le v_k\le25$~m/s,
$-4.0\le a_{x,k}\le2.2$~m/s$^2$, and
$|a_{y,k}|\le2.2$~m/s$^2$. For the motion-sickness cost in~\eqref{eq:misc_cost}, we use
$M_0=0.75$, $M_1=1$, $M_{\max}=10$, and $\alpha_M=0.75$.

We implement Problem~1 in C++ using the NLopt SLSQP solver. The
control interval is $\Delta t=0.25$~s and the prediction horizon is
$H=40$ steps ($10$~s). To reduce the optimization dimension, we use four piecewise-constant command blocks over the prediction horizon. Each learning and evaluation episode has a maximum duration of $60$~s
and terminates earlier if a physical collision is detected.

We set $\lambda_p=\lambda_s=0.5$ and initialize
$w_v=w_M=w_a=1/3$. The TTC threshold is $TTC_{th}=5$~s, and the
safety-supervisor horizon is $H_S=6$ steps ($1.5$~s).

\subsection{Simulated Passenger Profiles}

We consider three synthetic passenger profiles, Alice, Bob, and Zoey,
designed to represent a speed-oriented preference, a continuous
speed--comfort tradeoff, and a threshold-sensitive response to motion
sickness, respectively. These profiles are test cases rather than
population-level models of passenger behavior.

Each profile is evaluated under low, medium, and high motion-sickness
susceptibility, yielding nine scenarios. We select the corresponding
AM2 parameters from the individual parameter clusters reported by
Kotian et al.~\cite{Kotian2025}, as summarized in
Table~\ref{tab:am2_parameters}.

\begin{table}[t]
\vspace*{5pt}
\centering
\caption{Motion-sickness susceptibility parameters.}
\label{tab:am2_parameters}
\begin{tabular}{c|cc}
\hline
Susceptibility & $\theta_K$ & $\theta_T$ (s)\\
\hline
Low    & 9.5  & 86.2\\
Medium & 10.8 & 27.2\\
High   & 41.5 & 58.4\\
\hline
\end{tabular}
\vspace{-5pt}
\end{table}

For passenger $p$, let $e_{k+1}^{p}$ denote the instantaneous emotion
score and $g_k^{p}$ the vision mode.

Alice is the speed-oriented reference profile:
\begin{equation}
e_{k+1}^{\mathrm{Alice}}
=
|v_{k+1}|,
\qquad
g_k^{\mathrm{Alice}}
=
\mathrm{internal}.
\label{eq:alice_profile}
\end{equation}
Thus, susceptibility affects her predicted MISC state but not her
preference signal.

Bob represents a continuous tradeoff among speed, acceleration, and
motion sickness:
\begin{equation}
\begin{aligned}
e_{k+1}^{\mathrm{Bob}}
={}&
|v_{k+1}|
-2|a_{x,k}|\\
&-3|a_{y,k}|
-4M_{k+1}.
\end{aligned}
\label{eq:bob_emotion}
\end{equation}
His vision mode changes with MISC according to
\begin{equation}
g_k^{\mathrm{Bob}}
=
\begin{cases}
\mathrm{internal}, & M_k<0.8,\\
\mathrm{external}, & 0.8\le M_k<2,\\
\mathrm{closed},   & M_k\ge2.
\end{cases}
\label{eq:bob_vision}
\end{equation}

Zoey represents a threshold-sensitive passenger. Define
\[
\rho(v)=
\operatorname{clip}
\left(
\frac{|v|}{25},0,1
\right).
\]
Her emotion signal is
\begin{equation}
e_{k+1}^{\mathrm{Zoey}}
=
\begin{cases}
\rho(v_{k+1}),
& M_{k+1}<1,\\
-1000\,\operatorname{clip}(M_{k+1},1,10),
& M_{k+1}\ge1.
\end{cases}
\label{eq:zoey_emotion}
\end{equation}
and her vision mode is
\begin{equation}
g_k^{\mathrm{Zoey}}
=
\begin{cases}
\mathrm{internal}, & M_k<1,\\
\mathrm{external}, & 1\le M_k<2,\\
\mathrm{closed},   & M_k\ge2.
\end{cases}
\label{eq:zoey_vision}
\end{equation}
The discontinuity in~\eqref{eq:zoey_emotion} is intentional: it tests
whether the learned smooth weighted objective can accommodate a
preference that changes sharply.

\subsection{Learning and Evaluation Protocol}

For each of the nine scenarios, we collect $256$ learning episodes and
update the preference vector every $L=8$ episodes, yielding $32$
updates per scenario. Each posterior update uses four MCMC chains with
$1000$ burn-in and $4000$ sampling steps per chain. We retain every
second sample, yielding $2000$ retained samples per chain, and use a
proposal standard deviation of $0.3$.

For evaluation, we use four fixed-seed episodes per scenario that are
excluded from learning. We evaluate the initial uniform weights and
the weights after each of the $32$ updates using the same seeds. The
full-method study therefore contains $2304$ learning episodes and
$1188$ evaluation episodes.

\section{Closed-Loop Evaluation and Analysis}

We evaluate the learned controller by comparing the initial uniform
preference vector with the final learned preference vector using the
same four fixed evaluation seeds for each passenger--susceptibility
scenario. We report mean speed $\bar v$, mean MISC score $\bar M$,
root-mean-square acceleration $a_{\mathrm{RMS}}$, and mean emotion.
The mean speed and MISC score are trapezoidal time averages that
include the initial state. We define
\begin{equation}
a_{\mathrm{RMS}}
=
\sqrt{
\frac{1}{N}
\sum_k
\left(
a_{x,k}^2+a_{y,k}^2
\right)
},
\label{eq:arms}
\end{equation}
while the mean emotion is computed over the $N=240$ executed control
steps of each episode.

\subsection{Learned Passenger Preferences}

We first examine how the learned preference weights vary across
passenger profiles and susceptibility levels.
Table~\ref{tab:final_weights} reports the final preference weights. The results show distinct learned preference vectors across
the three passenger profiles and susceptibility settings.

For Alice, the learned preference vector becomes strongly
speed-oriented in all three susceptibility settings, with $w_v\in[0.865,0.983]$.
This behavior is consistent with the construction of Alice's emotion
signal in~\eqref{eq:alice_profile}, which depends only on vehicle
speed. The learned controller consequently assigns only a
small fraction of the preference weight to motion sickness and
acceleration.

For Bob, the learned weights exhibit a different pattern. In the low-
and medium-susceptibility cases, the largest weight is assigned to
motion sickness, whereas in the high-susceptibility case most of the
weight is distributed between speed and acceleration. These weights
should not be interpreted as estimates of the coefficients in
\eqref{eq:bob_emotion}. The emotion model uses linear speed and
absolute acceleration, whereas the controller uses a squared speed
deficit, normalized squared acceleration, and the nonlinear
motion-sickness cost in~\eqref{eq:misc_cost}. In addition,
acceleration influences both the instantaneous acceleration cost and
the subsequent evolution of motion sickness. Thus, the learned weights
represent the relative contribution of the controller features in
explaining the observed episode comparisons rather than a direct
recovery of the coefficients in Bob's simulated emotion function.

For Zoey, increasing susceptibility changes the learned allocation
primarily from speed toward acceleration. In particular, $w_v$
decreases from $0.874$ in the low-susceptibility case to $0.610$ in
the high-susceptibility case, while $w_a$ increases from $0.032$ to
$0.331$. The explicit weight assigned to the MISC cost remains
relatively small. This result illustrates an important feature of the
learning problem: because vehicle acceleration also affects the
subsequent MISC state, correlated trajectory features can produce
different weight allocations that explain similar episode-level
preferences.

\begin{table}[t]
\vspace*{5pt}
\centering
\caption{Final learned passenger preference weights.}
\label{tab:final_weights}
\begin{tabular}{ccccc}
\hline
Passenger & Susceptibility & $w_v$ & $w_M$ & $w_a$\\
\hline
Alice & Low    & 0.865 & 0.098 & 0.037\\
Alice & Medium & 0.983 & 0.003 & 0.014\\
Alice & High   & 0.973 & 0.003 & 0.024\\
\hline
Bob & Low    & 0.135 & 0.806 & 0.059\\
Bob & Medium & 0.200 & 0.648 & 0.152\\
Bob & High   & 0.550 & 0.027 & 0.423\\
\hline
Zoey & Low    & 0.874 & 0.093 & 0.032\\
Zoey & Medium & 0.823 & 0.087 & 0.090\\
Zoey & High   & 0.610 & 0.059 & 0.331\\
\hline
\end{tabular}
\vspace{-5pt}
\end{table}

The weight trajectories also exhibit empirical stabilization over the
$32$ preference updates. Across the last three updates, the largest
scenario-wise mean $\ell_1$ change in the preference vector is $0.030$
for Bob under high susceptibility. We emphasize that this observation
indicates limited late-stage variation in the reported runs rather
than convergence of the learning procedure in a theoretical sense.
A post-hoc four-chain diagnostic run using the same posterior and proposal settings yields a worst-case
$\widehat{R}=1.0048$ and a minimum effective sample size of $492$,
providing additional evidence that the posterior sampling procedure
mixed adequately in the reported experiments.

\subsection{Closed-Loop Performance}

We next examine how the learned preferences change closed-loop
behavior relative to the initial uniform weights.
Table~\ref{tab:performance} compares the initial and final controllers
using the fixed evaluation seeds. The mean emotion improves in seven
of the nine passenger--susceptibility scenarios. However, the seed-level results show that an improvement in the four-episode mean does not necessarily imply improvement in every evaluation episode.

\begin{table*}[t]
\vspace*{5pt}
\centering
\caption{Fixed-seed evaluation results with the initial and final
preference weights, averaged over four evaluation episodes per
scenario. Each entry reports initial $\rightarrow$ final.}
\label{tab:performance}
\begin{tabular}{cccccc}
\hline
Passenger & Susceptibility &
$\bar v$ (m/s) &
$\bar M$ &
$a_{\mathrm{RMS}}$ (m/s$^2$) &
Mean emotion\\
\hline
Alice & Low
& 17.34 $\rightarrow$ 18.81
& 0.244 $\rightarrow$ 0.388
& 0.361 $\rightarrow$ 0.679
& 17.343 $\rightarrow$ 18.815\\
Alice & Medium
& 19.82 $\rightarrow$ 20.07
& 0.472 $\rightarrow$ 0.735
& 0.407 $\rightarrow$ 0.728
& 19.822 $\rightarrow$ 20.071\\
Alice & High
& 17.94 $\rightarrow$ 19.83
& 0.553 $\rightarrow$ 1.136
& 0.355 $\rightarrow$ 0.848
& 17.942 $\rightarrow$ 19.839\\
\hline
Bob & Low
& 17.19 $\rightarrow$ 17.16
& 0.254 $\rightarrow$ 0.229
& 0.402 $\rightarrow$ 0.390
& 15.433 $\rightarrow$ 15.486\\
Bob & Medium
& 18.41 $\rightarrow$ 17.17
& 0.449 $\rightarrow$ 0.365
& 0.361 $\rightarrow$ 0.319
& 15.846 $\rightarrow$ 15.095\\
Bob & High
& 16.48 $\rightarrow$ 17.22
& 0.510 $\rightarrow$ 0.594
& 0.346 $\rightarrow$ 0.378
& 13.781 $\rightarrow$ 14.089\\
\hline
Zoey & Low
& 17.31 $\rightarrow$ 19.02
& 0.252 $\rightarrow$ 0.398
& 0.346 $\rightarrow$ 0.755
& 0.693 $\rightarrow$ 0.761\\
Zoey & Medium
& 18.66 $\rightarrow$ 18.89
& 0.466 $\rightarrow$ 0.595
& 0.343 $\rightarrow$ 0.566
& 0.747 $\rightarrow$ 0.756\\
Zoey & High
& 18.64 $\rightarrow$ 18.89
& 0.559 $\rightarrow$ 0.647
& 0.327 $\rightarrow$ 0.386
& 0.746 $\rightarrow$ -33.250\\
\hline
\end{tabular}
\vspace{-5pt}
\end{table*}

For Alice, the final controller increases the mean speed in all three
susceptibility settings. This increase is accompanied by higher mean
MISC scores and RMS acceleration, which follows directly from the
speed-oriented preference signal in~\eqref{eq:alice_profile}. For
example, in the low-susceptibility case, the mean speed increases from
$17.34$ to $18.81$~m/s, while the mean MISC score increases from
$0.244$ to $0.388$. The seed-level results also reveal substantial
variability: for Alice-low and Alice-medium, only one of the four
evaluation seeds improves in emotion, but the corresponding increase
is sufficiently large to yield a positive four-seed mean.

For Bob-low, the learned controller reduces the mean MISC score from
$0.254$ to $0.229$ while maintaining approximately the same mean
speed, resulting in a small increase in mean emotion. For Bob-high,
the controller accepts increases in both MISC and acceleration in
exchange for a higher mean speed, and the mean emotion increases from
$13.781$ to $14.089$. Bob-medium exhibits the opposite outcome: speed,
MISC, and RMS acceleration all decrease, but mean emotion decreases
from $15.846$ to $15.095$. In this case, one evaluation seed incurs an
emotion decrease of approximately $4.3$, which outweighs the
improvements observed in the other three seeds.

For Zoey-low and Zoey-medium, the learned controller increases both
mean speed and mean emotion while the MISC state remains below the
noticeable-symptom threshold in the evaluation episodes. The
high-susceptibility case exposes a limitation of the learned smooth
weighted objective. Although the mean speed increases from $18.64$ to
$18.89$~m/s, one of the four evaluation episodes crosses the
threshold $M=1$, reaching a maximum MISC value of $1.079$. During
this episode, $M\ge1$ for $7.75$~s, which activates the large penalty
in~\eqref{eq:zoey_emotion} and produces a mean emotion of $-135.207$.
Consequently, the four-seed mean emotion decreases from $0.746$ to
$-33.250$. The maximum MISC values in the other three final
evaluation episodes are $0.833$, $0.864$, and $0.750$.

This result highlights the distinction between learning a smooth
weighted surrogate objective and satisfying a threshold-sensitive
passenger preference. The current preference-learning mechanism does
not impose $M<1$ as a constraint. Therefore, a learned weight vector
that performs well over many episodes can still fail to prevent an
individual threshold crossing.

\subsection{Safety and Constraint Satisfaction}

We then examine whether safety constraints are preserved. 

No physical collision is observed in the $2304$ learning episodes or
the $1188$ evaluation episodes of the full method. The safety
supervisor intervenes on $1.621\%$ of all learning control steps. For
the final learned controllers, the mean intervention rate across the
four evaluation episodes ranges from $0.625\%$ to $4.271\%$, as
reported in Table~\ref{tab:safety_results}.

\begin{table}[t]
\centering
\caption{Safety and motion-sickness statistics under the final learned
weights, averaged over four evaluation episodes per scenario.
$M_{\max}$ and $TTC_{\min}$ are averages of the corresponding
per-episode maximum and minimum values.}
\label{tab:safety_results}
\begin{tabular}{ccccc}
\hline
Passenger & Susceptibility &
$M_{\max}$ & $TTC_{\min}$ (s) & Override (\%)\\
\hline
Alice & Low    & 0.554 & 2.734 & 2.292\\
Alice & Medium & 0.973 & 3.460 & 1.354\\
Alice & High   & 1.589 & 1.550 & 4.271\\
Bob   & Low    & 0.373 & 4.496 & 1.250\\
Bob   & Medium & 0.558 & 4.093 & 1.146\\
Bob   & High   & 0.843 & 3.811 & 0.625\\
Zoey  & Low    & 0.573 & 3.081 & 1.979\\
Zoey  & Medium & 0.799 & 2.959 & 1.875\\
Zoey  & High   & 0.881 & 3.347 & 1.771\\
\hline
\end{tabular}
\end{table}

The supervisor reaches the best-effort branch in
\eqref{eq:safety_supervisor}, where no admissible candidate is found
in the finite candidate set, on $34$ learning control steps and $15$
evaluation control steps. Thus, the absence of observed collisions
should be interpreted as an empirical result of these simulations and
not as a formal guarantee of collision avoidance.

The logged TTC is calculated from the instantaneous relative velocity
and collision ellipse at the pre-control state, whereas the planner
evaluates collision risk along its predicted trajectory. Therefore,
the reported minimum TTC can fall below the $5$~s penalty threshold
without violating a hard TTC constraint, since Problem~1 does not
impose such a constraint. Among the final-weight evaluations,
Alice-high has the smallest mean episode-minimum TTC of $1.550$~s.

The simulations also expose a limitation of the current fallback and
supervisor implementation. The nominal vehicle-motion limits are
exceeded on $86$ learning control steps and $20$ evaluation control
steps. Of these $106$ occurrences, $90$ arise during numerical
planner fallback and $16$ during safety-supervisor intervention. The
largest recorded lateral acceleration magnitude is
$11.706$~m/s$^2$, compared with the nominal limit of
$2.2$~m/s$^2$. Thus, satisfaction of the constraints in Problem~1 by
the nominal optimized trajectory does not guarantee their satisfaction
after a fallback or supervisory override. Enforcing the same motion
constraints within these two mechanisms is therefore necessary before
physical implementation.

\subsection{Feature Ablation}

We next investigate whether the acceleration and motion-sickness
features provide distinct information to the preference-learning
procedure. We repeat the complete $256$-episode learning protocol and
fixed-seed evaluation with two reduced preference models. The VA
variant retains only speed and acceleration, while the VM variant
retains only speed and MISC. In each reduced model, the two preference
weights are initialized uniformly at $0.5$. The full method retains
all three preference features.

\begin{table}[t]
\centering
\caption{Feature-ablation results under the final learned weights.
$a_{\mathrm{RMS}}$ is averaged over the nine scenario means.
Zoey failures count evaluation episodes with $M_{\max}\ge1$ among
the $12$ Zoey evaluations.}
\label{tab:ablation}
\begin{tabular}{ccccc}
\hline
Method &
$a_{\mathrm{RMS}}$ &
\shortstack{\rule{0pt}{2.0ex}Zoey-mid\\emotion} &
\shortstack{\rule{0pt}{2.0ex}Zoey\\failures} &
\shortstack{\rule{0pt}{2.0ex}Highest\\mean emotion}\\
\hline
VA   & 0.554 & -167.517 & 2/12 & 4/9\\
VM   & 0.719 & 0.762    & 0/12 & 4/9\\
Full & 0.561 & 0.756    & 1/12 & 1/9\\
\hline
\end{tabular}
\end{table}

Removing the acceleration feature in the VM variant increases the
aggregate RMS acceleration to $0.719$~m/s$^2$, compared with
$0.561$~m/s$^2$ for the full model and $0.554$~m/s$^2$ for VA. This
result indicates that the explicit acceleration feature influences
the smoothness of the resulting closed-loop trajectories.

Removing the MISC feature in the VA variant results in two
motion-sickness threshold crossings among the $12$ Zoey evaluation
episodes, compared with one crossing for the full model and none for
VM. Interestingly, VA avoids threshold crossing in all four
Zoey-high evaluation episodes. Thus, acceleration regulation can
indirectly reduce motion-sickness accumulation in some realizations,
but the results do not support treating acceleration as a consistent
replacement for the explicit MISC feature.

The ablation results also reveal a limitation of the current learning
formulation. The full three-feature model achieves the highest
four-seed mean emotion in only one of the nine scenarios, while VA
and VM each do so in four scenarios. Therefore, the results do not
show that adding all three features necessarily improves the learned
preference objective. Rather, they show that acceleration and MISC
have distinct effects on closed-loop behavior. Since acceleration
also influences subsequent motion sickness, the corresponding
features are correlated over an episode, and different preference
weight allocations can explain similar pairwise comparisons. More
data, richer preference feedback, or additional structure in the
preference model may be needed to improve identifiability.

All final-weight evaluation episodes for both reduced models are
collision-free. The VM variant recorded one collision among its
$2304$ learning episodes.

\section{Concluding Remarks}

In this paper, we presented a preference-adaptive receding-horizon
control framework that combines individualized motion-sickness
prediction with emotion-derived Bayesian preference learning and
independent safety supervision. The learned weights produce distinct
passenger-dependent closed-loop behaviors across the nine simulated
passenger--susceptibility scenarios, with mean evaluation emotion
improving in seven scenarios and no collisions observed for the full
method.

The results also expose limitations of the current formulation:
threshold-sensitive preferences can be poorly represented by a
weighted objective, and fallback or safety-supervisor interventions
can violate nominal motion constraints. Future work will investigate
human-subject feedback, uncertainty in motion-sickness estimation, and
preference models that better represent threshold-dependent and
time-varying passenger responses.

\section*{Disclosure of Large Language Model Use}

OpenAI ChatGPT~\cite{OpenAIChatGPT} and Codex~\cite{OpenAICodex}
were used to support literature search, coding,
and language editing. The authors reviewed and validated all material
incorporated into the manuscript and take full responsibility for the
content of the paper.

\bibliographystyle{IEEEtran}
\bibliography{current_references,IDS_Publications_07122026}

@inproceedings{wang2025visiopath,
	arxivid = {2507.06441},
	author = {Wang, Shanting and Typaldos, Panagiotis and Li, Chenjun and Malikopoulos, Andreas A},
	booktitle = {IEEE Open Journal of Control Systems (to appear)},
	title = {VisioPath: Vision-Language Enhanced Model Predictive Control for Safe Autonomous Navigation in Mixed Traffic},
	year = {2025}}

@article{le2025controller,
	arxivid = {arXiv:2311.00126},
	author = {Le, Viet-Anh and Malikopoulos, Andreas A},
	journal = {IEEE Robotics and Automation Letters},
	number = {8},
	pages = {8308-8315},
	publisher = {IEEE},
	title = {{Controller Adaptation via Learning Solutions of Contextual Bayesian Optimization}},
	volume = {10},
	year = {2025}}

@inproceedings{Le2023ACC,
	arxivid = {arXiv:2210.00700},
	author = {Le, Viet-Anh and Malikopoulos, Andreas A},
	booktitle = {2023 American Control Conference (ACC)},
	organization = {IEEE},
	pages = {1183--1188},
	title = {Optimal weight adaptation of model predictive control for connected and automated vehicles in mixed traffic with Bayesian optimization},
	year = {2023}}

@inproceedings{ChalakiCBF2022,
	author = {Chalaki, Behdad and Malikopoulos, Andreas A},
	booktitle = {Proceedings of the 61th IEEE Conference on Decision and Control (CDC)},
	pages = {2264-2269},
	title = {A Barrier-Certified Optimal Coordination Framework for Connected and Automated Vehicles},
	year = {2022}}

@inproceedings{Marceddu2021,
  author    = {A. C. Marceddu and J. Sini and M. Violante and B. Montrucchio},
  title     = {A novel approach to improve the social acceptance of autonomous driving vehicles by recognizing the emotions of passengers},
  booktitle = {Proc. SPIE 11605, 13th International Conference on Machine Vision},
  year      = {2021},
  pages     = {116051R},
  doi       = {10.1117/12.2586417}
}

@inproceedings{Zhao2022,
  author    = {Z. Zhao and Z. Wang and K. Han and R. Gupta and P. Tiwari and G. Wu and M. J. Barth},
  title     = {Personalized car following for autonomous driving with inverse reinforcement learning},
  booktitle = {Proc. 2022 IEEE International Conference on Robotics and Automation (ICRA)},
  year      = {2022},
  pages     = {2891--2897},
  doi       = {10.1109/ICRA46639.2022.9812446}
}

@article{Wen2024,
  author  = {X. Wen and X. Zheng and Z. Cui and S. Jian and D. He},
  title   = {Preference-based reinforcement learning for autonomous vehicle control considering the benefits of following vehicles},
  journal = {IEEE Transactions on Intelligent Vehicles},
  volume  = {11},
  number  = {8},
  pages   = {882--893},
  year    = {2026},
  doi     = {10.1109/TIV.2024.3412977}
}

@inproceedings{Ellis2024,
  author    = {E. Ellis and G. R. Ghosal and S. J. Russell and A. D. Dragan and E. B{\i}y{\i}k},
  title     = {A generalized acquisition function for preference-based reward learning},
  booktitle = {Proc. 2024 IEEE International Conference on Robotics and Automation (ICRA)},
  year      = {2024},
  pages     = {2814--2821},
  doi       = {10.1109/ICRA57147.2024.10611472}
}

@inproceedings{Wada2020,
  author    = {T. Wada and J. Kawano and Y. Okafuji and A. Takamatsu and M. Makita},
  title     = {A computational model of motion sickness considering visual and vestibular information},
  booktitle = {Proc. 2020 IEEE International Conference on Systems, Man, and Cybernetics (SMC)},
  year      = {2020},
  pages     = {1758--1763},
  doi       = {10.1109/SMC42975.2020.9283350}
}

@inproceedings{Kotian2023,
  author    = {V. Kotian and D. M. Pool and R. Happee},
  title     = {Modelling individual motion sickness accumulation in vehicles and driving simulators},
  booktitle = {Proc. 22nd Driving Simulation \& Virtual Reality Conference (DSC Europe 2023)},
  year      = {2023},
  pages     = {71--79}
}

@article{Kotian2024,
  author  = {V. Kotian and T. Irmak and D. M. Pool and R. Happee},
  title   = {The role of vision in sensory integration models for predicting motion perception and sickness},
  journal = {Experimental Brain Research},
  year    = {2024},
  volume  = {242},
  number  = {3},
  pages   = {685--725},
  doi     = {10.1007/s00221-023-06747-x}
}

@article{Kotian2025,
  author  = {V. Kotian and D. M. Pool and R. Happee},
  title   = {Personalizing motion sickness models: Estimation and statistical modeling of individual-specific parameters},
  journal = {Frontiers in Systems Neuroscience},
  year    = {2025},
  volume  = {19},
  pages   = {1531795},
  doi     = {10.3389/fnsys.2025.1531795}
}

@article{Wada2021,
  author  = {T. Wada},
  title   = {Computational model of motion sickness describing the effects of learning exogenous motion dynamics},
  journal = {Frontiers in Systems Neuroscience},
  year    = {2021},
  volume  = {15},
  pages   = {634604},
  doi     = {10.3389/fnsys.2021.634604}
}

@article{Tian2025,
  author  = {Z. S. Tian and S. E. Zhao and X. Xiao and H. B. Wei and K. Wang and J. Zeng},
  title   = {Evaluation of intelligent electric vehicle carsickness performance considering passenger motion sensitivity},
  journal = {Proceedings of the Institution of Mechanical Engineers, Part D: Journal of Automobile Engineering},
  year    = {2025},
  volume  = {239},
  number  = {14},
  pages   = {7076--7089},
  doi     = {10.1177/09544070241309693}
}

@article{Emond2026,
  author  = {W. Emond and R. Sauerbier and U. Scholly and F. Sasangohar and M. Zare},
  title   = {Motion sickness detection and mitigation in a stop-and-go passenger ride scenario},
  journal = {Transportation Research Part F: Traffic Psychology and Behaviour},
  year    = {2026},
  volume  = {116},
  pages   = {103381},
  doi     = {10.1016/j.trf.2025.103381}
}

@article{Tang2026,
  author  = {B. Tang and B. Luo and Y. Wu and Q. Tang},
  title   = {Research on a motion sickness prediction model for vehicle occupants based on vehicle dynamics parameter thresholds},
  journal = {Applied Ergonomics},
  year    = {2026},
  volume  = {135},
  pages   = {104774},
  doi     = {10.1016/j.apergo.2026.104774}
}

@inproceedings{kong2015,
  author    = {J. Kong and M. Pfeiffer and G. Schildbach and F. Borrelli},
  title     = {Kinematic and dynamic vehicle models for autonomous driving control design},
  booktitle = {Proc. IEEE Intelligent Vehicles Symposium (IV)},
  year      = {2015},
  pages     = {1094--1099},
  doi       = {10.1109/IVS.2015.7225830}
}

@article{daza2023,
  author  = {I. G. Daza and R. Izquierdo and L. M. Mart{\'{i}}nez and O. Benderius and D. Fern{\'{a}}ndez Llorca},
  title   = {Sim-to-real transfer and reality gap modeling in model predictive control for autonomous driving},
  journal = {Applied Intelligence},
  year    = {2023},
  volume  = {53},
  number  = {10},
  pages   = {12719--12735},
  doi     = {10.1007/s10489-022-04148-1}
}

@article{Oman1990,
  author  = {C. M. Oman},
  title   = {Motion sickness: a synthesis and evaluation of the sensory conflict theory},
  journal = {Canadian Journal of Physiology and Pharmacology},
  year    = {1990},
  volume  = {68},
  number  = {2},
  pages   = {294--303},
  month   = feb,
  doi     = {10.1139/y90-044}
}

@article{Schwarz2014,
  author  = {C. Schwarz},
  title   = {On computing time-to-collision for automation scenarios},
  journal = {Transportation Research Part F: Traffic Psychology and Behaviour},
  year    = {2014},
  volume  = {27},
  pages   = {283--294},
  doi     = {10.1016/j.trf.2014.06.015}
}

@article{Hou2014,
  author  = {J. Hou and G. F. List and X. Guo},
  title   = {New algorithms for computing the time-to-collision in freeway traffic simulation models},
  journal = {Computational Intelligence and Neuroscience},
  year    = {2014},
  volume  = {2014},
  pages   = {761047},
  doi     = {10.1155/2014/761047}
}

@misc{OpenAIChatGPT,
  author       = {{OpenAI}},
  title        = {{ChatGPT}},
  year         = {2022},
  note         = {[Online]. Available: https://openai.com/index/chatgpt/}
}

@misc{OpenAICodex,
  author       = {{OpenAI}},
  title        = {{Codex}},
  year         = {2026},
  note         = {[Online]. Available: https://openai.com/codex/}
}

\end{document}